\documentclass[final,5p,times,twocolumn]{elsarticle}

\usepackage{amssymb}

\usepackage[brazil]{babel}

\usepackage{times}
\usepackage{graphicx}
\usepackage{dcolumn}
\usepackage{bm}
\usepackage{hyperref}
\usepackage[usenames,dvipsnames]{color}
\usepackage{amsmath}
\usepackage{amsfonts}
\usepackage{amsthm}
\usepackage{mathrsfs}
\usepackage{subfigure}

\usepackage[normalem]{ulem}

\usepackage[utf8]{inputenc}

\def\ket#1{\mathinner{|{#1}\rangle}}

\DeclareMathOperator{\tr}{Tr}

\def\ket#1{\mathinner{|{#1}\rangle}}

\journal{Physics Letters A}

\begin{document}

\begin{frontmatter}



\title{Tripartite State Characterization via Activated Bipartite Entanglement}


\author[inst1,inst2]{L. G. E. Arruda}
\affiliation[inst1]{organization={Departamento de Engenharia de Computacao e Sistemas Digitais -  Universidade de Sao Paulo},
            addressline={Av. Prof. Luciano Gualberto, trav.3, 158},
            city={Sao Paulo},
            postcode={05508-010},
            state={SP}, 
            country={Brazil}}

\affiliation[inst2]{organization={QuaTI - Quantum Technology \& Information},
            addressline={Rua Major Jose Inacio},
            city={Sao Carlos},
            postcode={13560-161},
            state={SP}, 
            country={Brazil}}            

\author[inst3,inst4]{W. F. Balthazar}

\affiliation[inst3]{organization={Instituto Federal de Educacao, Ciencia e Tecnologia do Rio de Janeiro - IFRJ},
            addressline={Address One},
            city={Volta Redonda},
            postcode={00000-000},
            state={RJ}, 
            country={Brazil}}
\affiliation[inst4]{organization={Programa de Pos-graduacao em Fisica, Instituto de Fisica, Universidade Federal Fluminense},
            addressline={Address Two},
            city={Niteroi},
            postcode={00000-000},
            state={RJ}, 
            \country={Brazil}}
            
\author[inst5]{M. V. Moreira}
\affiliation[inst5]{organization={Instituto de Fisica Gleb Wataghin, Universidade de Campinas},
            addressline={Address Three}, 
            city={Campinas},
            postcode={13083-970}, 
            state={SP},
            country={Brazil}}

\author[inst6]{M. H. M. Passos}
\affiliation[inst6]{organization={ICFO—Institut de Ciencies Fotoniques, the Barcelona Institute of Science and Technology},
            addressline={Address Four}, 
            city={Castelldefels (Barcelona)},
            postcode={08860}, 
            country={Spain}}
            
\author[inst7]{J. A. O. Huguenin}
\affiliation[inst7]{organization={Instituto de Ciencias Exatas, Universidade Federal Fluminense},
            addressline={Address Five}, 
            city={Volta Redonda},
            postcode={00000-000},
            state={RJ}, 
            country={Brazil}}

\author[inst5]{M. C. de Oliveira}

\begin{abstract}
We propose a procedure to identify and classify genuine tripartite entanglement in pure 3-qubit states via the Activated Bipartite Entanglement (ABE), which is defined here as the difference between the Entanglement of Assistance and the Entanglement of Formation. We show that for pure states belonging to one of the two inequivalent classes of genuine tripartite entanglement, i.e., GHZ or W states, the ABE is always greater than zero. For separable and biseparable states it is always null. In addition, our approach is capable to distinguish between genuine tripartite entangled states, those belonging to the GHZ-class from those belonging to the W-class. We also present an experimental proposal, by using linear optical circuits and internal degrees of freedom of a single photon, to measure the ABE and to verify the characterization via activated entanglement. The circuit simulation shows an excellent agreement with theoretical prediction for a wide class of GHZ and W states.
\end{abstract}




\end{frontmatter}


\section{Introduction}

After decades of theoretical and experimental research, entanglement has gained the status of fundamental non-local resource, necessary to accomplish informational tasks in a more efficient way than in classical systems. While bipartite entanglement is well understood and properly quantified for arbitrarily mixed states, for both discrete \cite{Steinhoff2010} and continuous variable \cite{mco1}, quantitative and qualitative characterization of multipartite entanglement has been challenging \cite{mco2}.

In this manuscript, we propose a procedure to quantitatively distinguish pure 3-qubit states possessing genuine tripartite entanglement from biseparable and separable states by means of what we called the Activated Bipartite Entanglement (ABE), which quantifies the amount of entanglement shared by the combined system $ABC$ that can be exclusively localised in the subsystem $AB$ by local operations on subsystem $C$. The ABE is defined as the difference between two dual entanglement measures: the Entanglement of Assistance (EoA) \cite{EOA} and the Entanglement of Formation (EoF) \cite{wootters1998entanglement}. Moreover, it allows to distinguish among the genuine tripartite entangled states, those belonging to the GHZ class from those belonging to the W class. This difference between the posterior and the anterior amount of entanglement localised in $AB$ by means of local operations on $C$, also enables a clear ordering of states in a parameter map. Through the Generalized Schmidt Decomposition \cite{acin2000generalized}, for states with one independent parameter (1-parameter GHZ-type) and two independent parameters (2-parameter W-type) \cite{acin2001classification}, we obtain exact analytical results, shedding light on how pure tripartite states are distributed in a parametric space defined by three parametric functions given by the ABE, its upper bound and the dependent Schmidt coefficient. We also perform an numerical analysis considering more
general states.

Furthermore, we present an experimental proposal for the generation of 1-parameter GHZ and 2-parameter W states employing linear optics. Linear optical circuits are very precise and an excellent test bed for entanglement analysis \cite{multiphoton, environ1, environ2, environ3}. Besides, internal degrees of freedom of twin-photons were explored for study of multipartite states \cite{GHZOAM}. They have been used, for instance, for the implementation of the sudden death entanglement induced by environments emulated by linear transformations of polarization of twin-photons \cite{environ1}, by polarization mode dispersion \cite{environ2} and by an all-optical local CNOT \cite{environ3}. Internal degrees of freedom of twin-photons were also explored for preparation and geometry study of GHZ states \cite{GHZOAM, 18qubits}. By using path, polarization, and transverse mode degrees of freedom of single photons, it was previously proposed a full linear optical circuit to prepare tripartite GHZ state \cite{Balthazar16a, Balthazar16b}. Here, also using these three degrees of freedom we propose an experimental detection of ABE for 1-parameter GHZ and 2-parameter W states.

\section{Entanglement of Assistance}
\label{sec_EoA}

The definition of the EoA \cite{EOA} is motivated by the situation in which three parties spatially separated, $ABC$ ($A$, $B$ and $C$ stand for Alice, Bob and Charlie, respectively), share many copies of a pure tripartite entangled state given by $\rho_{ABC} = \vert\psi_{ABC}\rangle\langle\psi_{ABC}\vert$, and $AB$ would like to use their subsystems to perform some particular task. However, the reduced state $\rho_{AB} = \tr_{C}\left[\rho_{ABC}\right]$ might not be very pure nor sufficiently entangled for this purpose. However, $C$ can effectively help them by remotely transforming the available resource through appropriate local projective measurements in his subsystem. This procedure concentrates the initial tripartite entanglement in a new set of bipartite copies shared by $AB$. The more accurately chosen is the measurement basis, the higher is the concentrated entanglement. The rate that $AB$ can convert their copies (regardless of $C$'s assistance) into singlets to perform their task is given by \cite{bennett1996concentrating}
\begin{equation}
    \Bar{S}\left(\varepsilon\right) = \sum_{i}p_{i}S\left(\rho^{i}_{A}\right) = - \sum_{i}p_{i}\tr\left(\rho^{i}_{A}\log\rho^{i}_{A}\right) \label{conversion_rate},
\end{equation}
where $\varepsilon = \left\{p_{i},\rho^{i}_{AB}\right\}$ represents the ensemble made of pure states $\rho^{i}_{AB} = \vert\psi^{i}_{AB}\rangle\langle\psi^{i}_{AB}\vert$ with classical probabilities $p_{i}$. The EoA is defined as
\begin{equation}
    E_{A}\left(\rho_ {AB}\right)= \max_{\varepsilon}\Bar{S}\left(\varepsilon\right) = \max_{\left\{p_{i},\rho^{i}_{AB}\right\}}\sum_{i}p_{i}S\left(\rho^{i}_{A}\right), \label{EoA}
\end{equation}
i.e., it is the maximization of the conversion rate given by Eq.~\eqref{conversion_rate} over all convex (spectral) decomposition of $\rho_{AB}$ into pure states, that is, over all ensembles $\varepsilon$ for which $\rho_{AB} = \sum_{i}p_{i}\vert\psi^{i}_{AB}\rangle\langle\psi^{i}_{AB}\vert$. The EoA \cite{EOA} is recognised as the dual measure of the EoF in the sense that it is defined as the maximum value of the average entropy $\Bar{S}\left(\varepsilon\right)$ over all possible pure state decomposition of $\rho_{AB}$, while the EoF is defined as the minimum value of the same quantity \cite{wootters1998entanglement},
\begin{equation}
    E_{F}\left(\rho_ {AB}\right) = \min_{\varepsilon}\Bar{S}\left(\varepsilon\right) = \min_{\left\{p_{i},\rho^{i}_{AB}\right\}}\sum_{i}p_{i}S\left(\rho^{i}_{A}\right). \label{EoF}
\end{equation}

Note that the EoF captures the amount of entanglement shared by the pair $AB$ independently from action on subsystem $C$. On the other hand, the EoA captures the total amount of entanglement that is shared by $AB$ after the assistance of $C$. Nevertheless, the EoA does not distinguish whether $\rho_{AB}$ was previously entangled or not. 

To apply Eq.~\eqref{EoA} in some concrete examples, let us consider an arbitrary orthonormal basis to accomplish the projective measurements 
\begin{align}
\label{eqn:eqlabel}
\begin{split}
 \vert\xi\rangle& = \cos\theta\vert0\rangle_{C} + e^{i\phi}\sin\theta\vert1\rangle_{C},
\\
 \vert\chi\rangle& = \sin\theta\vert0\rangle_{C} - e^{i\phi}\cos\theta\vert1\rangle_{C}, 
\end{split}
\end{align}
where $\left\{\vert 0\rangle_{C},\vert 1\rangle_{C}\right\}$ is the Charlie's computational basis formed by the Pauli's $\sigma_{z}$ eigen-vectors. Then, the EoA and the EoF are, respectively, the maximum and the minimum values of a nonlinear function of two variables, $\Bar{S}\left(\theta,\phi\right)$,  
\begin{equation}
E_{A}\left(\theta,\phi\right ) = \max_{\left\{\theta,\phi\right\}}\Bar{S}\left(\theta,\phi\right) \label{obj_func_eoa},
\end{equation}
\begin{equation}
E_{F}\left(\theta,\phi\right ) = \min_{\left\{\theta,\phi\right\}}\Bar{S}\left(\theta,\phi\right) \label{obj_func_eof},
\end{equation}
where
\begin{equation}
   \Bar{S}\left(\theta,\phi\right) =  \sum_{i}p_{i}(\theta,\phi)S\left(\frac{\tr_{BC}\left[\Pi_{i}\rho_{ABC}\Pi_{i}\right]}{p_{i}(\theta,\phi)}\right), \label{averege_entropy}
\end{equation}
is the average entropy, $\Bar{S}\left(\theta,\phi\right) = p_{\xi}\left(\theta,\phi\right)S(\rho^{\xi}_{A}) + p_{\chi}\left(\theta,\phi\right)S(\rho^{\chi}_{A})$. The projectors in Eq.~\eqref{averege_entropy} are the two projectors of $C$ in the basis $\left\{\vert\chi\rangle, \vert\xi\rangle\right\}$, i.e., $\Pi_{i}\equiv \Pi_{i}(\theta,\phi) = \boldsymbol{1}_{A}\otimes\boldsymbol{1}_{B}\otimes\vert i\rangle\langle i\vert$ with $i=\xi,\chi$ (see Eq.~\eqref{eqn:eqlabel}), and $p_{i}(\theta,\phi) = \tr_{ABC}\left[\Pi_{i}(\theta,\phi)\rho_{ABC}\Pi_{i}(\theta,\phi)\right]$ is the associate probability distribution.

The optimization problem involved in Eqs.~\eqref{obj_func_eoa} and~\eqref{obj_func_eof} is not a trivial task, specially if the matrix that represents the tripartite pure state is a dense matrix. However, there are important cases where it is possible to calculate the solutions exactly. In this manuscript we present both: analytical solutions for these special types of states belonging to the GHZ and W classes, and numerical solutions for arbitrary pure tripartite states obtained from the Globally Convergent Method for Systems of Non-Linear Equations described in Ref.~\cite{press1986numerical}. 
Nonetheless, before we go through the analytical and numerical results, we present in the next section a quantity that demonstrates to be a helpful tool for distinguishing genuine tripartite entangled states from biseparable and separable tripartite states and, beyond, gives an upper bound to W states. 

\section{Activated Bipartite Entanglement} \label{The_Difference_of_Entanglement}

To account exclusively for the extra amount of entanglement that can be localised in the subsystem $AB$ through local projective measurements on $C$, the Activated Bipartite Entanglement (ABE) is defined as the maximal difference between the average entropy, $\Bar{S}\left(\varepsilon\right)$, of the reduced state $\rho_A$, anterior and posterior the action on $C$,
\begin{eqnarray}
    \Delta_{E}\left(\rho_ {AB}\right) &\equiv& 
\max_{\varepsilon}\left[\Bar{S}\left(\varepsilon_{\text{p}}\right)-
\Bar{S}\left(\varepsilon_{\text{a}}\right)\right]\nonumber \\
 &=&\max_{\varepsilon_{\text{p}}}\Bar{S}\left(\varepsilon_{\text{p}}\right)-\min_{\varepsilon_{\text{a}}}\Bar{S}\left(\varepsilon_{\text{a}}\right),\label{DoE_new}
\end{eqnarray}
where  $\varepsilon_{\text{a}}$ and $\varepsilon_{\text{p}}$ represent the anterior (before operation on $C$) and posterior (after operation on $C$) ensembles of pure state decomposition of $\rho_{AB} = \tr_{C}\rho_{ABC}$ and $\rho_{AB}^{\text{p}} \equiv \tr_{C}\left[\sum_i\Pi_{i}\rho_{ABC}\Pi_{i}\right]$. Therefore, the ABE given by Eq.~\eqref{DoE_new} can be expressed as the difference between EoA and EoF as given by Eqs.~\eqref{EoA} and \eqref{EoF},
\begin{equation}
    \Delta_{E}\left(\rho_ {AB}\right) = E_{A}\left(\rho_ {AB}\right)- E_{F}\left(\rho_ {AB}\right). \label{DoE}
\end{equation}
For many copies of a state $\rho_{ABC}$, the ABE quantifies the rate of maximally entangled bipartite states that are successfully activated. In other words, $E_{F}\left(\rho_ {AB}\right)$ measures the initial amount of entanglement between $AB$ that is already there before the assistance of $C$, while $\Delta_{E}\left(\rho_ {AB}\right)$ represents exclusively the extra amount of entanglement added to it after the assistance.

To conduct a proper analysis of the behavior of 
ABE for arbitrary pure tripartite states, let us consider its lower and upper bounds. First note that, since the EoA is the maximum value of the average entropy (given by Eq.~\eqref{conversion_rate}) over all pure state decomposition of $\rho_{AB}$ while EoF is the minimum value of the same function, by construction, the ABE cannot be negative and it is always greater than or equal to zero,
\begin{equation}
    \Delta_{E}\left(\rho_ {AB}\right) = E_{A}\left(\rho_ {AB}\right) - E_{F}\left(\rho_ {AB}\right) \geq 0. \label{DoE_lower_bound}
\end{equation}
Therefore $E_{F}\left(\rho_ {AB}\right)$ is the lower bound of $E_{A}\left(\rho_ {AB}\right)$, that is, $E_{F}\left(\rho_ {AB}\right)\leq E_{A}\left(\rho_ {AB}\right)$. As we will see in the next section, $\Delta_{E}\left(\rho_ {AB}\right) = 0$, i.e., $E_{A}\left(\rho_ {AB}\right)= E_{F}\left(\rho_ {AB}\right)$ only for biseparable and separable pure tripartite states. The physical meaning for ABE's vanishing values is the following: there is no local measurement that can be done by $C$ capable to concentrate any extra amount of entanglement between $AB$, besides the one that is already there before the measurement, if the combined system $ABC$ shares non-genuine tripartite entanglement. From Ref. \cite{EOA}, we get the upper bound,   
\begin{equation}
    \Delta_{E}\left(\rho_ {AB}\right) \leq
    \Delta S, \label{DoE_upper_bound}
\end{equation}
where
\begin{equation}
    \Delta S = S\left(\rho_{AB}\right) - \left\vert S\left(\rho_{A}\right) - S\left(\rho_{B}\right)\right\vert. \label{deltaS}
\end{equation}

\section{Biseparable and separable tripartite states}

 The fact that $\Delta_{E}\left(\rho_ {AB}\right) = 0$ for all biseparable and separable tripartite states can be easily checked, noting that $\Delta S$ in Eq.~\eqref{DoE_upper_bound} is always equals to zero for these kind of states. Let us show this point carrying out an analyses on a case by case basis.
 
 First, let us examine the trivial case of a pure tripartite separable state ($\rho_{ABC} = \rho_{A}\otimes\rho_{B}\otimes\rho_{C}$). In this case, both EoA and EoF are equal to zero because, in this particular case, for the combined state to be pure each part must be also pure which implies that $S\left(\rho_{A}\right) = S\left(\rho_{B}\right) = S\left(\rho_{C}\right) = S\left(\rho_{AB}\right) = 0$. Since $E_{F}\left(\rho_ {AB}\right)\leq E_{A}\left(\rho_ {AB}\right)\leq \min\left\{S\left(\rho_{A}\right), S\left(\rho_{B}\right)\right\}$ (see Ref.~\cite{EOA} for more details) and $E_{F}\left(\rho_ {AB}\right) = S\left(\rho_{A}\right)$ for the pure state $\rho_{AB}=\rho_{A}\otimes\rho_{B}$, we see that $E_{A}\left(\rho_{AB}\right)=E_{F}\left(\rho_{AB}\right)=0$. 
 
 Now, for biseparable states we have three distinct scenarios. The first one is the case where the part AB is not entangled with C ($\rho_{ABC} = \rho_{AB}\otimes\rho_{C}$). In this case, since the tripartite state $\rho_{ABC}$ is pure, the reduced states $\rho_{AB}$ and $\rho_{C} = \tr_{AB}\rho_{ABC}$ must be also pure, thereby $S\left(\rho_{AB}\right) = S\left(\rho_{C}\right) = 0$ and  $S\left(\rho_{A}\right) = S\left(\rho_{B}\right) = E_{F}\left(\rho_ {AB}\right)\rightarrow E_{F}\left(\rho_ {AB}\right)= E_{A}\left(\rho_ {AB}\right)$, considering that $E_{F}\left(\rho_ {AB}\right)\leq E_{A}\left(\rho_ {AB}\right)\leq \min\left\{S\left(\rho_{A}\right), S\left(\rho_{B}\right)\right\}$. In this case $0\leq S\left(\rho_{A}\right)\leq 1$, depending on the amount of entanglement between AB. The other two scenarios are the following: the system A is not entangled with BC ($\rho_{ABC} = \rho_{A}\otimes\rho_{BC}$) or the system B is not entangled with AC ($\rho_{ABC} = \rho_{AC}\otimes\rho_{B}$). In both cases, C is unable to help AB  because A and B are initially disentangled and there is no projective measurement that C can perform on his qubit that produces any entanglement between AB. Moreover, for the case where $\rho_{ABC} = \rho_{A}\otimes\rho_{BC}$, we have $S\left(\rho_{A}\right) = 0$, and for the case where $\rho_{ABC} = \rho_{AC}\otimes\rho_{B}$, we have $S\left(\rho_{B}\right) = 0$. In both cases $E_{A}\left(\rho_{AB}\right) = E_{F}\left(\rho_{AB}\right) = 0$, taking in to account that $\rho_{AB}$ is pure and $E_{F}\left(\rho_ {AB}\right)\leq E_{A}\left(\rho_ {AB}\right)\leq \min\left\{S\left(\rho_{A}\right), S\left(\rho_{B}\right)\right\}$.

\section{Exact results for the GHZ and W states}\label{Exact_optimization_procedure}

As mentioned at the end of Section~\ref{sec_EoA}, there are important cases where it is possible to calculate exactly the solutions of the problem posed by Eqs.~\eqref{obj_func_eoa} and~\eqref{obj_func_eof}. Fortunately, both the GHZ and W usual states are represented by sparse matrices and the average entropies given by Eq.~\eqref{averege_entropy} are simple functions of the  parameters that define the optimal basis given by Eqs.~\eqref{eqn:eqlabel}, even for more general states, namely the 1-parameter GHZ-type and the 2-parameter W-type that come from the Generalized Schmidt Decomposition \cite{acin2000generalized,acin2001classification}
\begin{equation}
\vert \psi\rangle = \lambda_{0}\vert 000\rangle + \lambda_{1}e^{i\varphi}\vert 100\rangle + \lambda_{2}\vert 101\rangle + \lambda_{3}\vert 110\rangle+ \lambda_{4}\vert 111\rangle, \label{acin}
\end{equation}
where $\lambda_{i}\geq 0$, $\sum_{i}\lambda^2_{i} = 1$ and $0\leq\varphi\leq\pi$. The state given by Eq.~\eqref{acin} describes arbitrary pure tripartite states with 5 independent parameters. Here, we calculate analytically the solutions of Eqs.~\eqref{obj_func_eoa},~\eqref{obj_func_eof} and~\eqref{averege_entropy} considering the 1-parameter GHZ-type~\eqref{GHZ_type_state1} and the 2-parameter W-type~\eqref{W_type_2_parameters1} as follows.

\subsection{1-parameter GHZ-states}

We start with the simplest case given by the well-known GHZ state
\begin{equation}
\label{eqn:GHZ_state}
\vert GHZ\rangle = \frac{1}{\sqrt{2}}\left(\vert 000\rangle + \vert 111\rangle\right),
\end{equation}
which is a particular case of the 1-parameter GHZ-type given by Eq.~\eqref{GHZ_type_state1}, and whose average entropy given by Eq.~\eqref{averege_entropy} is obtained exactly and depends solely on $\theta$ as a binary entropy with $p = \cos^{2}\theta$,
\begin{equation}
\label{eqn:Binary_entropy_GHZ}
    \Bar{S}_{\vert GHZ\rangle}\left(\theta\right) = -p\log_{2}p - \left(1-p\right)\log_{2}\left(1 - p\right).
\end{equation}
The EoA for the GHZ state is the maximum value of Eq.~\eqref{eqn:Binary_entropy_GHZ} which is attained when $p=1/2$ at $\theta = \pi/4$:
\begin{equation}
    E_{A}\left(\rho_{AB}\right) = \Bar{S}_{\vert GHZ\rangle}\left(\pi/4\right) = 1;
\end{equation}
on the other hand, for $p=1$ or $p=0$ we have the minimum value of $\Bar{S}_{\vert GHZ\rangle}\left(\theta\right)$ which are  reached at $\theta = 0$ and $\theta = \pi/2$, respectively. The minimum value corresponds to the EoF for the GHZ state:
\begin{equation}
    E_{F}\left(\rho_{AB}\right) = \Bar{S}_{\vert GHZ\rangle}\left(0\right) = \Bar{S}_{\vert GHZ\rangle}\left(\pi/2\right) = 0
\end{equation}
Therefore, the optimal basis that maximizes the average entropy~\eqref{averege_entropy} for the GHZ state is spanned by the eigenvectors of Pauli's $\sigma_{x}$ operator, while the basis that minimizes it is formed by the eigenvectors of Pauli's $\sigma_{z}$ operator. Since the EoF is always zero for any bi-partition of the GHZ state \footnote{This is due to the fact that the GHZ state is very fragile under particle losses, implying that if we trace out one of the three qubits that forms it the remaining bipartite state is completely unentangled.}, the  ABE is equal to the EoA: 
\begin{equation}
    \Delta_{E}\left(\rho_{AB}\right) = E_{A}\left(\rho_{AB}\right) =1. 
\end{equation}
In  addition, the GHZ state saturates the upper bound given by Eq.~\eqref{DoE_upper_bound}
\begin{equation}
\label{eqn:Saturation_GHZ}
    \Delta_{E}\left(\rho_{AB}\right) = S\left(\rho_{AB}\right),
\end{equation}
since $S\left(\rho_{AB}\right) = S\left(\rho_{A}\right) = S\left(\rho_{B}\right) = S\left(\rho_{C}\right) = 1$. This can be seen as a characteristic trait of these kind specific GHZ states.

Indeed, every 1-parameter GHZ state \cite{acin2000generalized,acin2001classification} given by 
\begin{equation}
    \ket{\Psi_{GHZ}}=\lambda_1 \ket{000} + \lambda_2 \ket{111},
    \label{GHZ_type_state1}
\end{equation}
with $\lambda_1$ and $\lambda_2\geq 0$ and $\lambda_{2} = \sqrt{1 - \lambda_{1}^2}$, presents the same features described above for the special case of GHZ state given by Eq.~\eqref{eqn:GHZ_state}. For instance, the EoF of any bi-partition of these states are always equal to zero. This imply that the ABE and the EoA are always equal to each other, $\Delta_{E}\left(\rho_{AB}\right) = E_{A}\left(\rho_{AB}\right)$, expressing the fact that if the threesome ABC shares a 1-parameter GHZ state given by Eq.~\eqref{GHZ_type_state1}, all the entanglement that AB can get to eventually perform some task comes necessarily from the C's assistance. Furthermore, for these kind of states
\begin{eqnarray}
    S\left(\rho_{AB}\right) &=& S\left(\rho_{A}\right) = S\left(\rho_{B}\right) = S\left(\rho_{C}\right) \nonumber \\
    &=&-\lambda^{2}_{1}\log_{2}\left(\lambda^{2}_{1}\right)-\lambda^{2}_{2}\log_{2}\left(\lambda^{2}_{2}\right)
\end{eqnarray}
(a result that can be easily checked) which implies that the right hand side of Eq.~\eqref{deltaS} reduces to $S\left(\rho_{AB}\right)$. Besides, the density matrix that represents the state given by Eq.~\eqref{GHZ_type_state1} is a sparse matrix which allowed us to calculate the corresponded average entropy~\eqref{averege_entropy} analytically. The result can be seen bellow and shows us that the average entropy is a simple function of the parameters that describe both, the 1 parameter GHZ state given by Eq.~\eqref{GHZ_type_state1} and the C's measurement basis formed by the vectors~\eqref{eqn:eqlabel}, :
\begin{eqnarray}
     \Bar{S}_{\vert \psi_{GHZ}\rangle}\left(\theta\right) = &-&\lambda_2^2\cos ^2(\theta)\log_2 \left(\frac{-2\lambda_2^2\cos ^2(\theta)}{\left(\lambda_1^2-\lambda_2^2\right) \cos (2 \theta)-1}\right)\nonumber \\
     &-&\lambda_1^2\cos ^2(\theta) \log_2 \left(\frac{2 \lambda_1^2 \cos ^2(\theta)}{\left(\lambda_1^2-\lambda_2^2\right) \cos (2 \theta)+1}\right)\nonumber \\
     &-&\lambda_1^2\sin ^2(\theta) \log_2 \left(\frac{-2 \lambda_1^2 \sin ^2(\theta)}{\left(\lambda_1^2-\lambda_2^2\right) \cos (2 \theta)-1}\right)\nonumber \\
     &-&\lambda_2^2\sin ^2(\theta) \log_2 \left(\frac{2\lambda_2^2 \sin ^2(\theta)}{\left(\lambda_1^2-\lambda_2^2\right) \cos (2 \theta)+1}\right).\label{aver_entropy_1parGHZ}\nonumber \\
     &&
\end{eqnarray}
The closed-form  of $\Bar{S}_{\vert \psi_{GHZ}\rangle}$ given by Eq.~\eqref{aver_entropy_1parGHZ} enables us to find the exact solutions for the optimization problem posed by Eqs.~\eqref{obj_func_eoa} and~\eqref{obj_func_eof}. The first thing to note is that the average entropy for 1-parameter GHZ states does not depend on the relative phase $\phi$ of C's measurement basis, depending solely on $\theta$: $\Bar{S}_{\vert \psi_{GHZ}\rangle}\left(\theta, \phi\right) = \Bar{S}_{\vert \psi_{GHZ}\rangle}\left(\theta\right)$. The second thing is that the maximum value is given by
\begin{equation}
     E_{A}\left(\rho_{AB}\right) = -\lambda^{2}_{1}\log_{2}\left(\lambda^{2}_{1}\right) -\lambda^{2}_{2}\log_{2}\left(\lambda^{2}_{2}\right),
\end{equation}
at $\theta = \pi/4$ and the minimum value, $E_{F} = 0$, at $\theta = 0$ and $\theta = \pi/2$. 

Therefore, we can conclude two things regarding the 1-parameter GHZ states: (i) it always saturate the bound ~\eqref{DoE_upper_bound} 

\begin{equation}
    \Delta_{E}\left(\rho_{AB}\right) = E_{A}\left(\rho_{AB}\right) = S\left(\rho_{AB}\right),
    \label{Saturation_upper_bound}
\end{equation}
showing that C's assistance transfer all the entanglement contained in the 1-parameter GHZ-state to AB. (ii) The bases that maximizes and minimizes the average entropy do not depend on $\lambda_1$: $\sigma_x$ basis always maximizes the 1-parameter GHZ states' average entropy while $\sigma_z$ basis always minimizes it.   

\begin{center}
\begin{figure}[h]
    \includegraphics[scale=0.3]{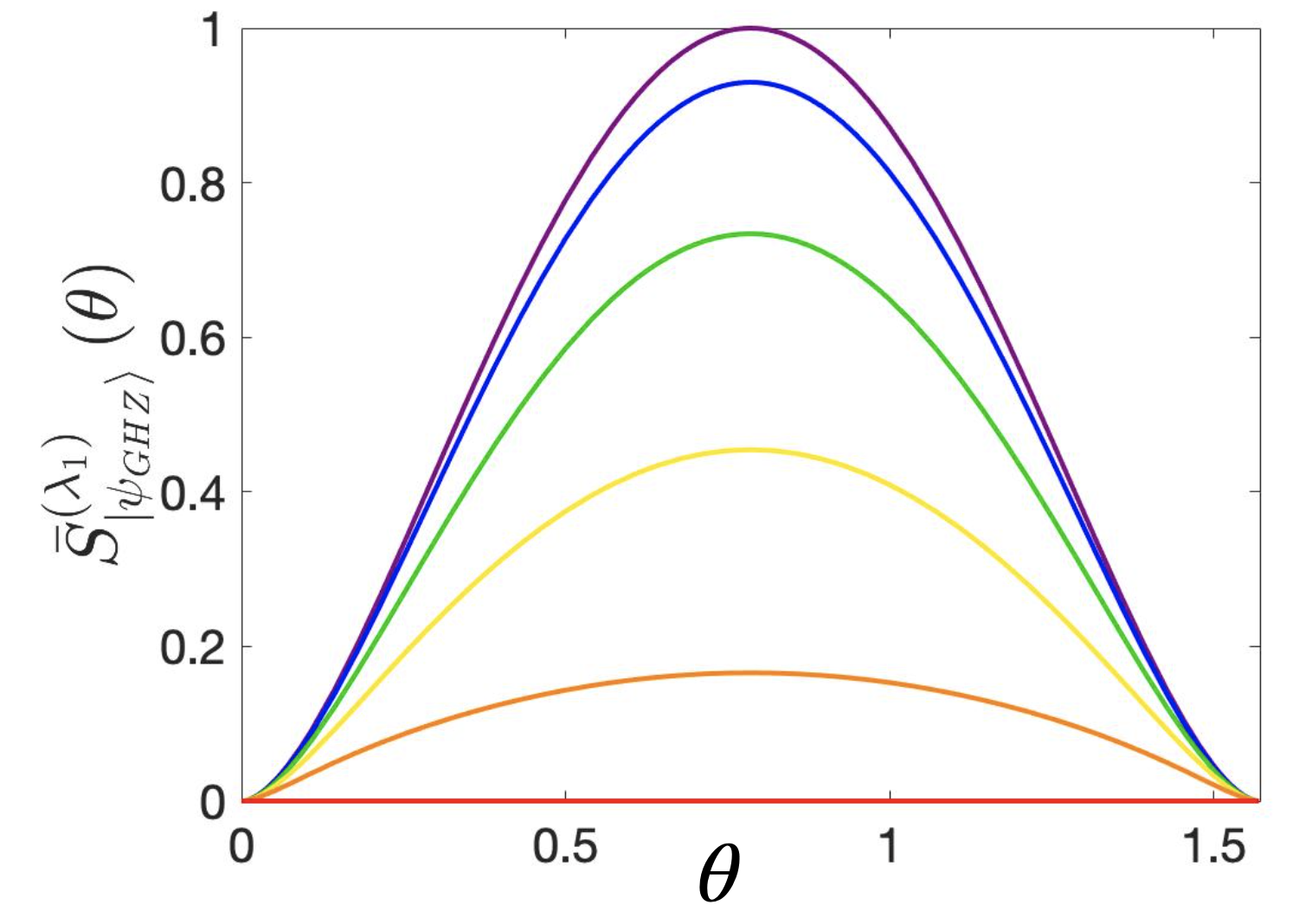}
    \caption{The average entropy given by Eq.~\eqref{averege_entropy} when $\rho_{ABC} = \vert\psi_{GHZ}\rangle\langle\psi_{GHZ}\vert$ for different values of $\lambda_{1}$ in Eq.~\eqref{GHZ_type_state1}. In red we have $\lambda_{1} = 1$ (or $\lambda_{1} = 0$). In orange $\lambda_{1} = 0.9877$. In yellow $\lambda_{1} = 0.9511$. In green $\lambda_{1} = 0.8910$. In blue $\lambda_{1} = 0.8090$. In purple we have the GHZ state given by Eq.~\eqref{eqn:GHZ_state}, i.e., $\lambda_{1} = \sqrt{2}/2 = 0.7071$.} 
    \label{fig:my_label2}
\end{figure}
\end{center}

Figure~\ref{fig:my_label2} shows the average entropy given by Eq.~\eqref{aver_entropy_1parGHZ} for different values of $\lambda_{1}$ varying $\theta$. In red we have $\lambda_{1} = 1$ (or $\lambda_{1} = 0$), in orange $\lambda_{1} = 0.9877$, in yellow $\lambda_{1} = 0.9511$, in green $\lambda_{1} = 0.8910$, in blue $\lambda_{1} = 0.8090$. In purple we have the GHZ state given by Eq.~\eqref{eqn:GHZ_state}, i.e., $\lambda_{1} = \sqrt{2}/2 = 0.7071$. Varying $\theta$ from $0$ to $\pi/2$, the maximal values that represent the EoA, and consequently the ABE itself, are always reached at $\theta = \pi/4$, which corresponds to measurements made in the $\sigma_{x}$ basis. The minimal values that represents the EoF are always reached at $\theta = 0$ or at $\theta = \pi/2$, which corresponds to measurements made in the $\sigma_{z}$ basis. The highest ABE is attained for $\lambda_{1} = \sqrt{2}/2 = 0.7071$, reducing its value for states that deviates from the uniform superposition found in the usual GHZ state, i.e, when $\lambda_1$ deviates from $\lambda_{1} = \sqrt{2}/2$.  

\subsection{2-parameter W-states}

We also solved the ABE analytically for the 2-parameter W-type given by 
\begin{equation}
    \vert \psi_{W2}\rangle = \lambda_{0}\vert 100\rangle + \lambda_{3}\vert 010\rangle + \lambda_{2}\vert 001\rangle, \label{W_type_2_parameters1}
\end{equation}
which is a particular case of W states given by 3 independent parameters (3-parameter W-type) \cite{acin2000generalized,acin2001classification}
\begin{equation}
\vert \psi_{W3}\rangle = \lambda_{0}\vert 000\rangle + \lambda_{1}\vert 100\rangle + \lambda_{2}\vert 101\rangle + \lambda_{3}\vert 110\rangle, \label{W_type_3_parameters1}
\end{equation}
where for both cases the coefficients must satisfy the constraints $\lambda_{i}\geq 0$ and $\sum_{i}\lambda^2_{i} = 1$. Note that we obtain the 2-parameter W-type~\eqref{W_type_2_parameters1} making $\lambda_{1} = 0$ and applying a BIT-FLIP operation on the first qubit of each product state of the remaining superposition in Eq.~\eqref{W_type_3_parameters1}.

Due to the sparsity of the density matrix that represents the 2-parameter W state~\eqref{W_type_2_parameters1}, it is possible to obtain the average entropy~\eqref{averege_entropy} analytically. The average entropy for this state depends solely on $\theta$, similar to the 1-parameter GHZ state. However, $\Bar{S}_{\vert W2\rangle}\left(\theta\right)$ is a convex function of the compact parameter $0\leq\theta\leq\pi/2$, with maximum value at $\theta = 0$ and $\theta = \pi/2$ and minimum value at $\theta = \pi/4$. Thus, the basis that maximizes~\eqref{averege_entropy} is formed by the eigenvectors of Pauli's $\sigma_{z}$ operator while the basis that minimizes~\eqref{averege_entropy} is formed by the eigenvectors of Pauli's $\sigma_{x}$ operator. This is in contrast to the 1-parameter GHZ state, where the bases that maximizes and minimizes the average entropy must be in opposite order of the 2-parameter W states. The EoF of W vectors given by Eqs.~\eqref{W_type_2_parameters1} and \eqref{W_type_3_parameters1} are not equal to zero in general, depending on the values of the independent parameters that define the state. This means that W states present entanglement between pairs AB, AC and BC before any assistance coming from a third part. We solved analytically both the EoF and EoA for the 2-parameter W states~\eqref{W_type_2_parameters1} and they are  given by the following expressions

\begin{eqnarray}
E_{F}\left(\rho^{W2}_ {AB}\right) &=& \mathcal{F}\left(\frac{1 - \sqrt{1 - 4\lambda_{3}^2\cdot\lambda_{0}^2}}{2}\right) \nonumber \\
&&+ \mathcal{F}\left(\frac{1 + \sqrt{1 - 4\lambda_{3}^2\cdot\lambda_{0}^2}}{2}\right) \label{EoF_W2_1},
\end{eqnarray}
\begin{equation}
E_{A}\left(\rho^{W2}_ {AB}\right) = -\mathcal{F}\left(\lambda_{0}^2 + \lambda_{3}^2\right) + \mathcal{F}\left(\lambda_{0}^2\right) + \mathcal{F}\left(\lambda_{3}^2\right), \label{EoA_W2_1}
\end{equation}
where $\mathcal{F}(x) = -x\log_{2}(x)$.
We also solved the upper bound $\Delta S^{W2}= S\left(\rho^{W2}_{AB}\right) - \left\vert S\left(\rho^{W2}_{A}\right) - S\left(\rho^{W2}_{B}\right)\right\vert$ of Eq.~\eqref{DoE_upper_bound} exactly: 
\begin{eqnarray}
\Delta S^{W2} &=& \mathcal{F}\left(\lambda_{0}^2 + \lambda_{3}^2\right) + \mathcal{F}\left(1 - \lambda_{0}^2 - \lambda_{3}^2\right)\nonumber \\
&&- \left\vert\mathcal{F}\left(\lambda_{0}^2\right) + \mathcal{F}\left(1 - \lambda_{0}^2\right) - \mathcal{F}\left(\lambda_{3}^2\right) - \mathcal{F}\left(1 - \lambda_{3}^2\right)\right\vert.\nonumber \\
&& \label{DeltaS_W2_1}
\end{eqnarray}

\begin{figure}[ht!]
    \includegraphics[scale=0.23]{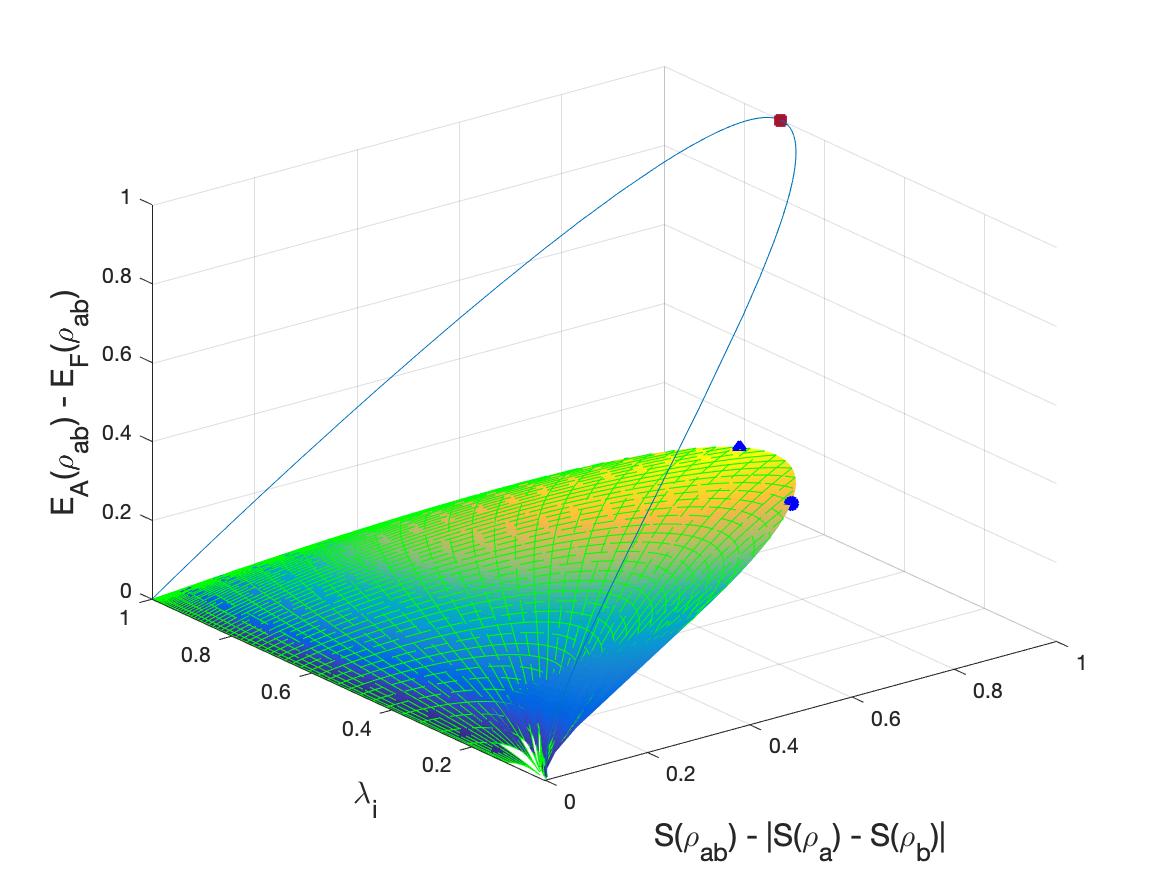}
    \caption{Parametric curve and parametric surface for both, 1-parameter GHZ and the 2-parameter W-type states, respectively. The parameter $\lambda_{i}$ is the dependent parameter. For 1-parameter GHZ states $\lambda_1^2 = 1 - \lambda_2^2$ and for 2-parameter W states $\lambda_2^2 = 1 - \lambda_0^2 - \lambda_3^2$. The usual GHZ and W states given by Eqs.~\eqref{eqn:GHZ_state} and ~\eqref{W_state} are also shown: respectively the red square with coordinates $(0.707, 1, 1)$ and the blue dot with coordinate $(0.5773, 0.9183, 0.11)$  }
    \label{fig:Surface}
\end{figure}

It is possible to visualize the results achieved so far trough a parametric space defined by coordinates $(x,y,z)= \left(\Delta S,\lambda_i,\Delta_{E}\left(\rho_ {AB}\right) \right)$.  We will call this space as the ABE's parametric space. From Eqs.~\eqref{EoF_W2_1} and \eqref{EoA_W2_1} we get $\Delta_{E}\left(\rho^{W2}_ {AB}\right) = E_{A}\left(\rho^{W2}_ {AB}\right) - E_{F}\left(\rho^{W2}_ {AB}\right)$ and together with $\Delta S^{W2}$ and the dependent parameter $\lambda_{2} = \sqrt{1 - \lambda_{0}^2 - \lambda_{3}^2}$, it is possible to see that the 2-parameter W states given by Eq.~\eqref{W_type_2_parameters1} live upon the parametric surface depicted in Fig.~\eqref{fig:Surface}. The usual W state, 
\begin{equation}
\vert W\rangle = \frac{1}{\sqrt{3}}\left(\vert 100\rangle + \vert 010\rangle + \vert 001\rangle\right) , \label{W_state}
\end{equation}
has $\Delta S^{W} = 0.9183$, $E_{A}\left(\rho^{W}_{AB}\right) = 0.666$ and $E_{F}\left(\rho^{W}_{AB}\right) = 0.55$, which means that $\Delta_{E}\left(\rho^{W}_{AB}\right) = 0.11$. Thus, the blue dot whose coordinates are $ \left(\Delta S^{W},\lambda_{2},\Delta_{E}\left(\rho^{W}_ {AB}\right)\right) = \left(0.9183,0.5773,0.11\right)$ is the usual W state representation on the ABE's parametric space depicted in Fig.~\eqref{fig:Surface}. Another important W state is the one among all W states with the highest ABE. The highest ABE attained by a W state is $\Delta_{E}\left(\rho^{W}_{AB}\right) = 0.1498$ and this state is depicted in Fig.~\eqref{fig:Surface} as a blue diamond by the following coordinates: $(0.7663, 0.9646, 0.1498)$. Note that the ABE for this state is $\Delta_{E}\left(\rho^{W}_{AB}\right) \approx 0.15$, which is far bellow the supposed upper bound $\Delta S^{W} = 0.9183$. This fact brings two key insights about how we can distinguish among the genuine tripartite entangled states, those belonging to the GHZ-class from those belonging to the W-class. States with higher ABE than $\Delta_{E}\left(\rho^{W}_{AB}\right) = 0.1498$ are necessarily GHZ states and the ABE's upper bound for W states is 
\begin{equation}
    \Delta_{E}\left(\rho^{W}_{AB}\right) = 0.1498.
\end{equation}

Also in Fig.~\eqref{fig:Surface}, it is possible to see the parametric curved line (above the 2-parameter W surface) contained on the plane $\Delta_{E}\left(\rho_{AB}\right) =  S\left(\rho_{AB}\right)$, which corresponds to the place where the 1-parameter GHZ states can be found. The usual GHZ state given by~\eqref{eqn:GHZ_state} is also plotted, the red square depicted by coordinates $\left(\Delta S^{GHZ},\lambda_{1},\Delta_{E}\left(\rho^{GHZ}_ {AB}\right)\right) = \left(1,0.707,1\right)$. We see that GHZ-class and W-class states live in antipodal regions of all the possible 3-qubit pure states. 

\section{Experimental proposal}

To observe the bounds for tripartite entanglement we must be able to prepare a wide class of 3-qubit GHZ and W states. Here we present an experimental proposal based on the utilization of internal degrees of freedom of light associated with single photon source produced by Spontaneous Parametric Down Conversion (SPDC) \cite{GHZOAM}.  In particular, we propose the use three degrees of freedom of a single photon to codify the qubits, namely the propagation path (p), the polarization (P), and the transverse modes (M). The tripartite state is described by $\ket{pPM}$. 
Experimental studies of GHZ states with the three degrees of freedom of light were performed in the regime of intense laser beams [14, 15]. By exploiting the classical-quantum analogy, Mermin's inequalities were violated  [14]. To the best of our knowledge, experimental implementation of tripartite states codified in internal degrees of freedom of a single photon has not yet been performed. In this section, we present a proposal and perform a numerical simulation of the circuit to show the accordance of ABE measurement for a physical system.

We codify subsystem $A$ in the polarization degree of freedom ($\ket{H} \equiv \ket{0}_A,\ket{V} \equiv \ket{1}_A$), subsystem $B$ in the first order of Hermitian-Gaussian modes ($\ket{HG_{10}}=\ket{h}\equiv \ket{0}_B, \ket{HG_{01}}=\ket{v}\equiv \ket{1}_B$), and subsystem $C$ in path degree of freedom ($\ket{up}=\ket{u}\equiv \ket{0}_C, \ket{down}=\ket{d}\equiv \ket{1}_C$) \cite{Balthazar16b}. We define the order of qubits as $\ket{C}\otimes \ket{A} \otimes \ket{B}$. i.e., a photon propagating in the path $u$, with polarization $H$, transverse mode $h$  has its tripartite state defined as $\ket{uHh} \equiv \ket{000}$, and, consequently, $\ket{dVv} \equiv \ket{111}$.
     
The preparation of tripartite states using the three internal degrees of freedom of a photon follows the proposal of Ref.~\cite{Balthazar16b}. Polarized single photons can be prepared with $h/v$ transverse modes and have their path controlled. Then, a general tripartite state can be prepared. Let us show the optical circuits and their simulation for GHZ and W states. \\ 
   
\subsection{1-parameter GHZ states}

\begin{figure}[!ht]
	\centering
	\includegraphics[scale=0.45,trim=0cm 0cm 0cm 0cm, clip=true]{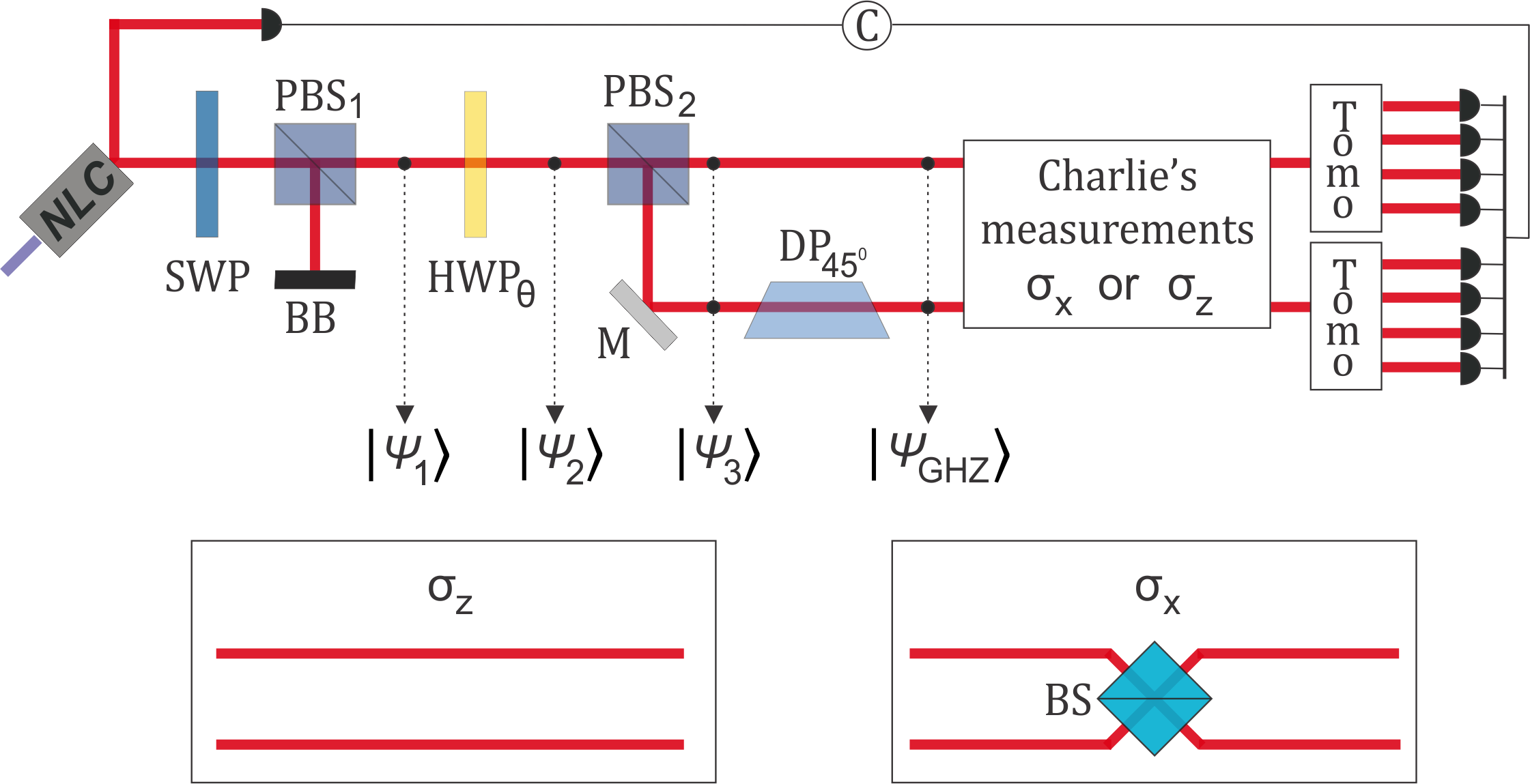}
  	\caption{Experimental setup for GHZ-type states. NLC: non linear crystal, SWP: S-wave plate to produce spin-orbit non-separable modes, PBS: Polarized Beam Splitter, HWP$_\theta$: half wave plate with its fast axis making an angle of $\theta$ with the horizontal plane, DP$_{45^\circ}$: Dove Prism with its base rotated $45^\circ$ with respect the horizontal plane, TOMO: tomography circuit for spin-orbit states as proposed in Ref.\cite{Balthazar21}. The tomography outputs in both paths are detected by photodetectors that are triggered by coincidence with the detection of Idler. BS: 50/50 non polarized Beam Splitter.}
	\label{setupGHZ}
\end{figure}

The schematic optical circuit for preparation of a single photon in a 1-parameter GHZ-state as defined in Eq.~\eqref{GHZ_type_state1}, is presented in Fig.~\ref{setupGHZ}. A SPDC source composed by a Laser of frequency $\omega_p$ pumping a Nonlinear Crystal (NLC) produces twin photons -- the Signal with frequency $\omega_s$ is used for GHZ preparation and the Idler with frequency $\omega_i$ is used to trigger the measurement apparatus. Naturally, the phase matching condition $\omega_p = \omega_s + \omega_i$ is satisfied.

The Signal passes through a S-Wave Plate (SWP) in order to prepare radial polarization, that corresponds to the spin-orbit entangled state of the system $AB$. In the first Polarized Beam Splitter (PBS$_1$), a polarization projective measurement is performed producing the state $\ket{\Psi_1}=\ket{Hh}\equiv \ket{00}_{AB}$. This is a simple and very efficient way to prepare the initial state.

 A path qubit can be prepared by a variable Beam Splitter that can be performed by the Half Wave Plate (HWP) with its fast axis making an angle of $\theta$ with the horizontal and a PBS. The HWP$_{\theta}$ transforms the polarization producing the following state 
\begin{equation}
    \ket{\Psi_2}= \ket{h} \otimes \left[ \cos(2\theta) \ket{H} + \sin (2\theta)\ket{V}\right].
\end{equation}
The PBS$_2$ projects H-polarization in path $u$ and V-polarization in path $d$. Then, we have a tripartite state prepared as 
\begin{equation}
    \ket{\Psi_3}=  \cos(2\theta) \ket{u H h } + \sin (2\theta)\ket{dVh}.
    \label{Psi3}
\end{equation}
Finally, the Dove Prism (DP) performs the transformation $\ket{h}\rightarrow \ket{v}$ of the transverse mode in the path $d$, and the Signal photon becomes the state
\begin{eqnarray}
    \ket{\Psi_{GHZ}}&=&  \cos(2\theta) \ket{u H h } + \sin (2\theta)\ket{dVv}\nonumber \\ &\equiv&\lambda_1 \ket{000} + \lambda_2 \ket{111},
    \label{GHZ_type_state}
\end{eqnarray}
where $\lambda_1 = \cos(2\theta)$ and $\lambda_2 = \sin(2\theta)$, with $\lambda_{2} = \sqrt{1 - \lambda_{1}^2}$. The preparation of 1-parameter GHZ state is complete. 

Let us discuss the measurement. At the end of each path we have a spin-orbit state tomography circuit as presented in \cite{Balthazar21}. With the tomography circuit it is possible to reconstruct the reduced density matrix $\rho_{AB}$. However, before the tomography $C$ needs to perform $\sigma_x$ and $\sigma_z$ measurements, since these are the bases that respectively optimize Eqs.~\eqref{EoA} and~\eqref{EoF} in the case of 1-parameter GHZ states. For $\sigma_z$, both paths are free and the click in the paths $u$ or $d$ gives the outcome 0 or 1 for this measure. For $\sigma_x$, path $u$ and $d$ are sent to a Beam Splitter (BS) \cite{Balthazar16a} and the output coming from it goes to tomography circuit. 

\begin{center}
\begin{figure}[h]
    \includegraphics[scale=0.27]{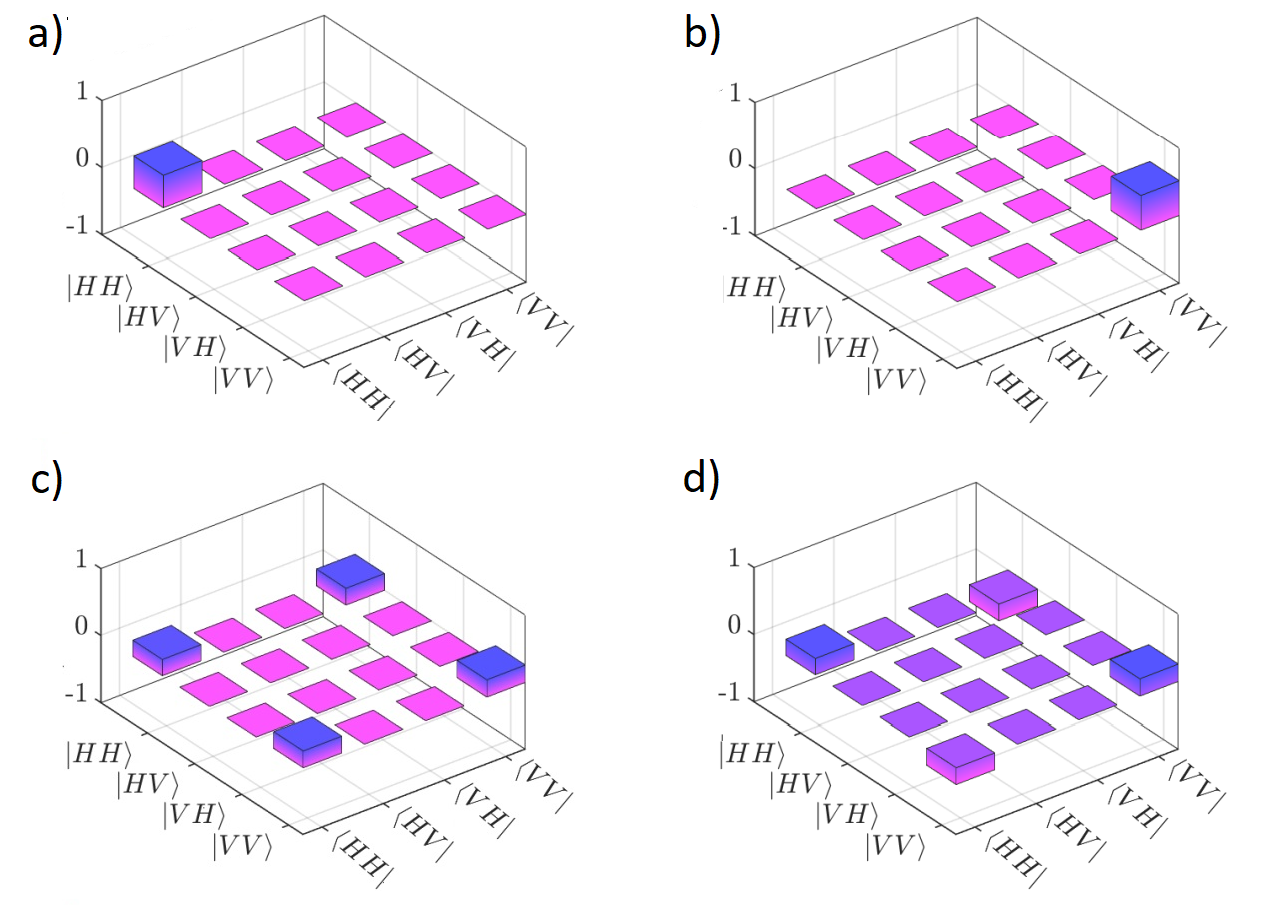}
    \caption{Results of the tomography for GHZ states after Charlie's measurement of a) $\sigma_z$ resulting 0, b) $\sigma_z$ resulting 1, c) $\sigma_x$ resulting 0, and d) $\sigma_x$ resulting 1. The  results are the density matrix of the reduced state $\rho_{AB}$ (spin-orbit state) multiplied by the corresponding weigh that depends on $\lambda_1$, $\lambda_2$, in this case $\frac{1}{\sqrt{2}}$.} 
    \label{rhoABGHZ}
\end{figure}
\end{center}

We simulate the preparation and measurement of 1-parameter GHZ-states by using Jones Matrix formalism \cite{Jones:41}. Our experimental proposal is designed to give as output the reduced density matrix $\rho_{AB}$ of spin-orbit states multiplied by the corresponding weight that depends on $\lambda_1$ and $\lambda_2$. The matrix obtained is used to calculate the difference: $E_A(\rho_{AB})- E_F(\rho_{AB})$. In order to emulate experimental errors, we introduce an error of $\pm 1^\circ$ in the angle $\theta$ of the HWP that defines the parameters $\lambda_1$ and $\lambda_2$.
Figure~\eqref{rhoABGHZ} presents the measurement results for $\lambda_1 = \lambda_2=\frac{1}{\sqrt{2}}$. Figure~\eqref{rhoABGHZ}-a) shows the obtained matrix for C's measurement in $\sigma_z$ resulting 0, Fig.~\eqref{rhoABGHZ}-b) resulting 1. Figure~\eqref{rhoABGHZ}-c) shows the results of $\rho_{AB}$ for Charlie's measurement of $\sigma_x$ resulting 0 and Fig.~\eqref{rhoABGHZ}-d) resulting 1.  The partial trace is obtained by the convex sum of the matrix of paths 0 and 1 with the respective weight. The obtained matrices by the optical circuit simulation are in excellent agreement with theoretical expectation presenting fidelity of 0.97, 0.99, 0.99, and 0.99, respectively. For this case we obtain $E_A(\rho_{AB})- E_F(\rho_{AB})=0.9992$, that presenting an error of less than $0.01\%$ with respect the theoretical expectation.    
%

\begin{figure}[h]
    \includegraphics[scale=0.23]{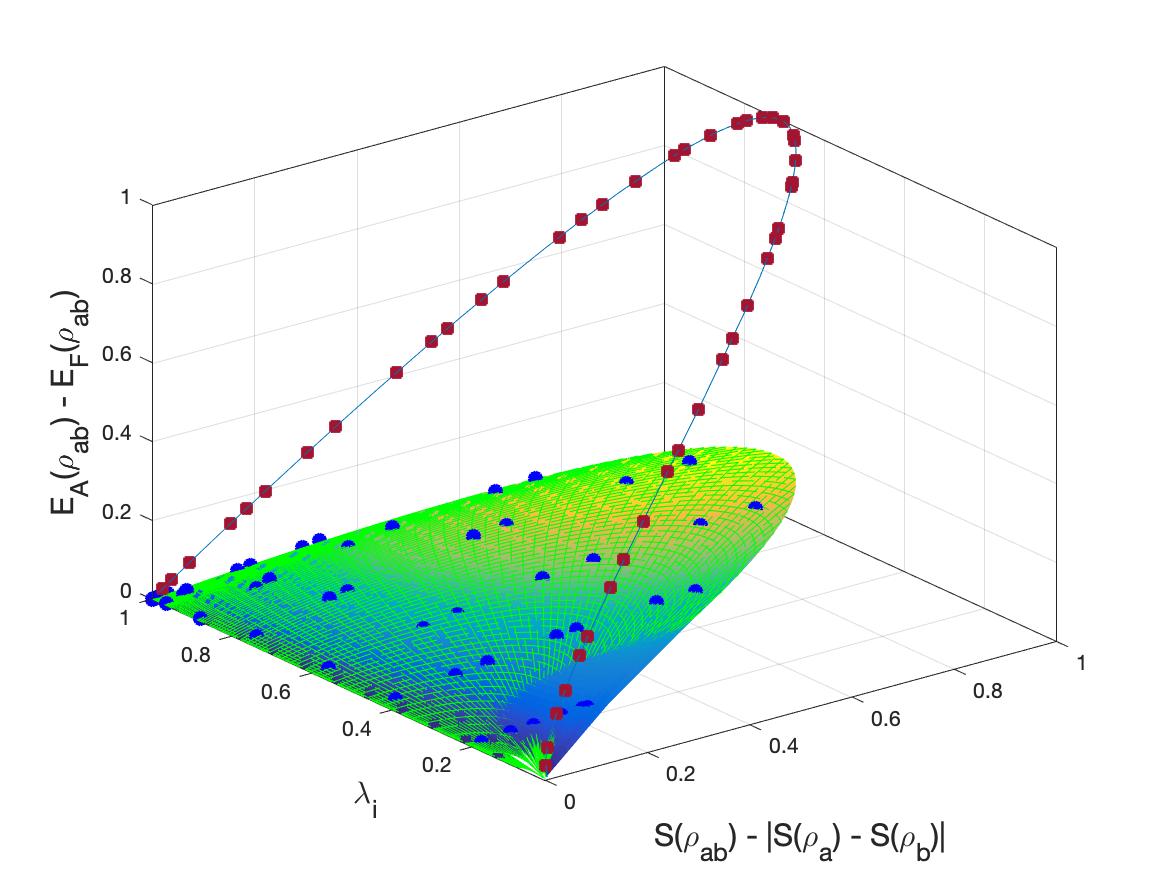}
    \includegraphics[scale=0.23]{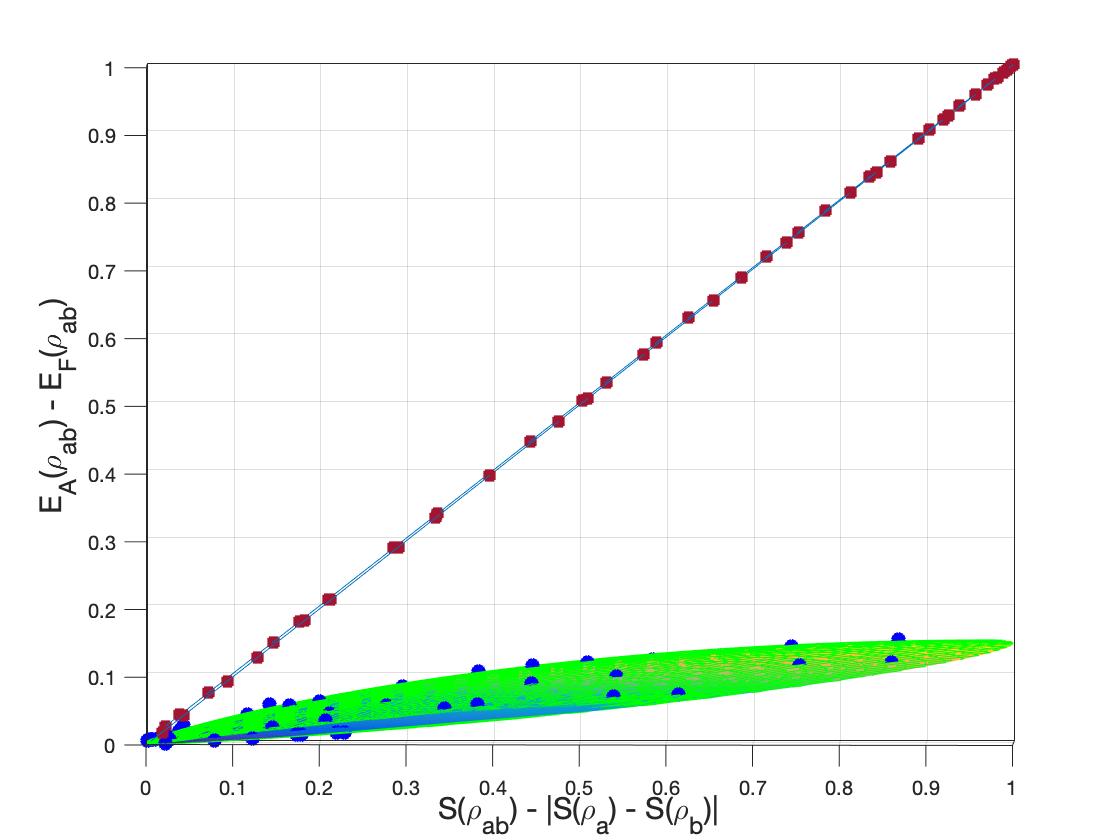}
    \caption{Top: Parametric curve and parametric surface for both, 1-parameter GHZ and 2-parameter W states, respectively. Bottom: $E_F(\rho_{AB}) \times S\left(\rho_{AB}\right) - \left\vert S\left(\rho_{A}\right) - S\left(\rho_{B}\right)\right\vert$. The red squares and the dark-blue dots are the outcomes stemming optical circuit simulation. The simulation considered typical $ \pm 1^\circ$ in the rotation element that fixes the parameters $\lambda_1$ and $\lambda_2$. The resulting error bar is neglected compared with the dots size.} 
    \label{fig:exp}
\end{figure}

Analytical results presented in Section \ref{Exact_optimization_procedure} show that 1-parameter GHZ-states saturates the upper bound given in Eq.~\eqref{DoE_upper_bound}, maximizing the amount of extra entanglement that can be localised in $AB$ via local operations on $C$. In Fig.~\ref{fig:exp}, the saturation given by Eq.~\eqref{Saturation_upper_bound} is represented in the ABE's parametric space. The blue curved line in Fig.~\ref{fig:exp}-top represents the 1-parameter GHZ states given by Eq.~\eqref{GHZ_type_state}. The red squares are the outcomes of the optical circuit simulation, whose results are in a remarkable agreement with theoretical expectation. The resulting error bar due the simulation of rotation of angle $\theta$ is neglected compared with the dots size. Irrespective of the value assumed by $\lambda_{1}$, the optimal measurement basis that maximizes the extra amount of entanglement between $AB$ is spanned by the eigenvectors of Pauli's $\sigma_{x}$ operator while the basis formed by the eigenvectors of Pauli's $\sigma_{z}$ is the basis that minimizes the $C$'s assistance, i.e., the one where $E_{A}\left(\rho_ {AB}\right) = E_{F}\left(\rho_ {AB}\right)$ and no extra entanglement is activated. Since the entanglement between any bi-partition is zero before operations on $C$ and, after that, is equal to the total amount of entanglement shared initially by the trio $ABC$, the assistance is a full assistance and the saturation given by Eq.~\eqref{Saturation_upper_bound} represents a characteristic only fulfilled by 1-parameter GHZ states.

\subsection{2-parameter W state}


\begin{figure}[!ht]
	\centering
	\includegraphics[scale=0.45,trim=0cm 0cm 0cm 0cm, clip=true]{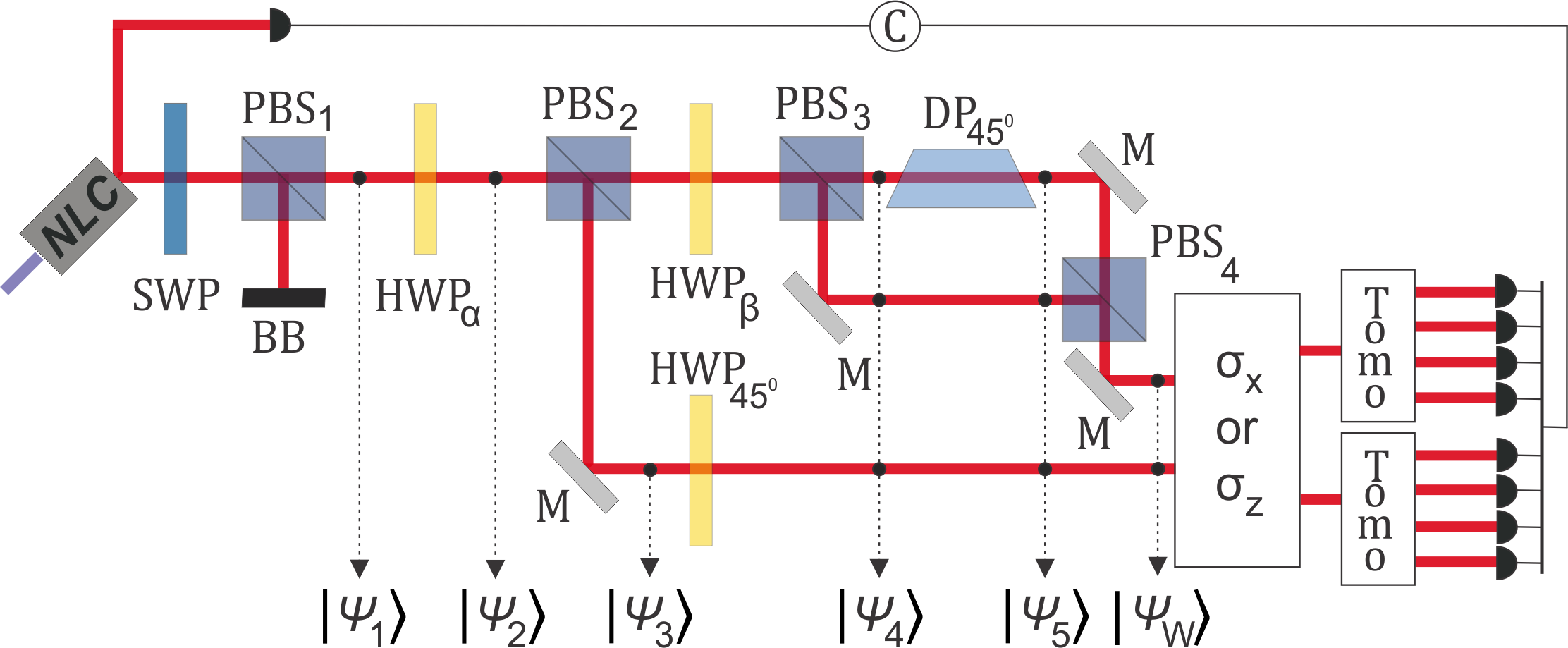}
	\caption{Experimental setup for W-type state preparation. NLC: non linear crystal, SWP: S-wave plate to produce spin-orbit nonseparablel modes, PBS: Polarized Beam Splitter, HWP$_\phi$: half wave plate with its fast axis making an angle of $\phi=\alpha,\  \beta,\  45^\circ$ with the horizontal plane, DP$_{45^\circ}$: Dove Prism with its base rotated $45^\circ$ with respect the horizontal plane, TOMO: tomography circuit for spin-orbit states as proposed in Ref.~\cite{Balthazar21}. The tomography outputs in both paths are detected by photocounters detectors that are triggered by coincidence with the detection of Idler.}
	\label{setupW}
\end{figure}

Figure~\ref{setupW} presents the schematic optical circuit for preparation and measurements of 2-parameter W states. The circuit is the same as for the preparation of 1-parameter GHZ states up to the point depicted in the Fig.~\ref{setupW} by the state $\ket{\Psi_3}$, which is given by Eq.~\eqref{Psi3}. Here, $\alpha$ is the angle of the first HWP for path preparation. A photon in the path $d$ passes through a HWP$_{45^\circ}$ and transforms its polarization as $\ket{V}\rightarrow \ket{H}$ and this part of the state is in accordance with the first term of Eq.~\eqref{W_type_2_parameters1}, $\lambda_0 \ket{100}$, where $\lambda_0= \sin(2\alpha)$ and $\ket{100}\equiv \ket{dHh}$. The other two terms are associated to path $u$, and we can  write $ \ket{0}_C \otimes\left( \lambda_3 \ket{10} + \lambda_2 \ket{01} \right)$. The terms in parentheses are an unbalanced entangled state in the spin-orbit degree of freedom of light with weights $\lambda_2$ and $\lambda_3$. It can be prepared by a C-NOT gate using polarization as the control qubit. The HWP$_\beta$ will transform 
\begin{equation}
    \cos(2\alpha)\ket{Hh} \rightarrow \cos(2\alpha)\left[ \cos(2\beta)\ket{H} + \sin(2\beta)\ket{V} \right]\otimes \ket{h}.
\end{equation}
The V-polarization is reflected by PBS$_3$ and is deviated towards PBS$_4$. The H-polarization is transmitted in PBS$_3$, passes through a DP that transforms the transverse mode as $\ket{h}\rightarrow\ket{v}$ and is transmitted by PBS$_4$ where is aligned with the other arm in the path $u$. Then, finally, we have
 \begin{eqnarray}
   \ket{\Psi_{W2} }&=&  \sin(2\alpha)\ket{dHh} +
   \cos(2\alpha)\cos(2\beta) \ket{uHv} \nonumber \\&& + \cos(2\alpha)\sin(2\beta)\ket{uVh}.
 \end{eqnarray}
Taking into account our codification and by comparing with Eq.~\eqref{W_type_2_parameters1}, we identify $\lambda_0 =\sin(2\alpha)$, $\lambda_2=\cos(2\alpha)\cos(2\beta)$, and $\lambda_3=\cos(2\alpha)\sin(2\beta)$. Here, the preparation of 2 parameter W state is complete. 

\begin{center}
\begin{figure}[h]
\includegraphics[scale=0.27]{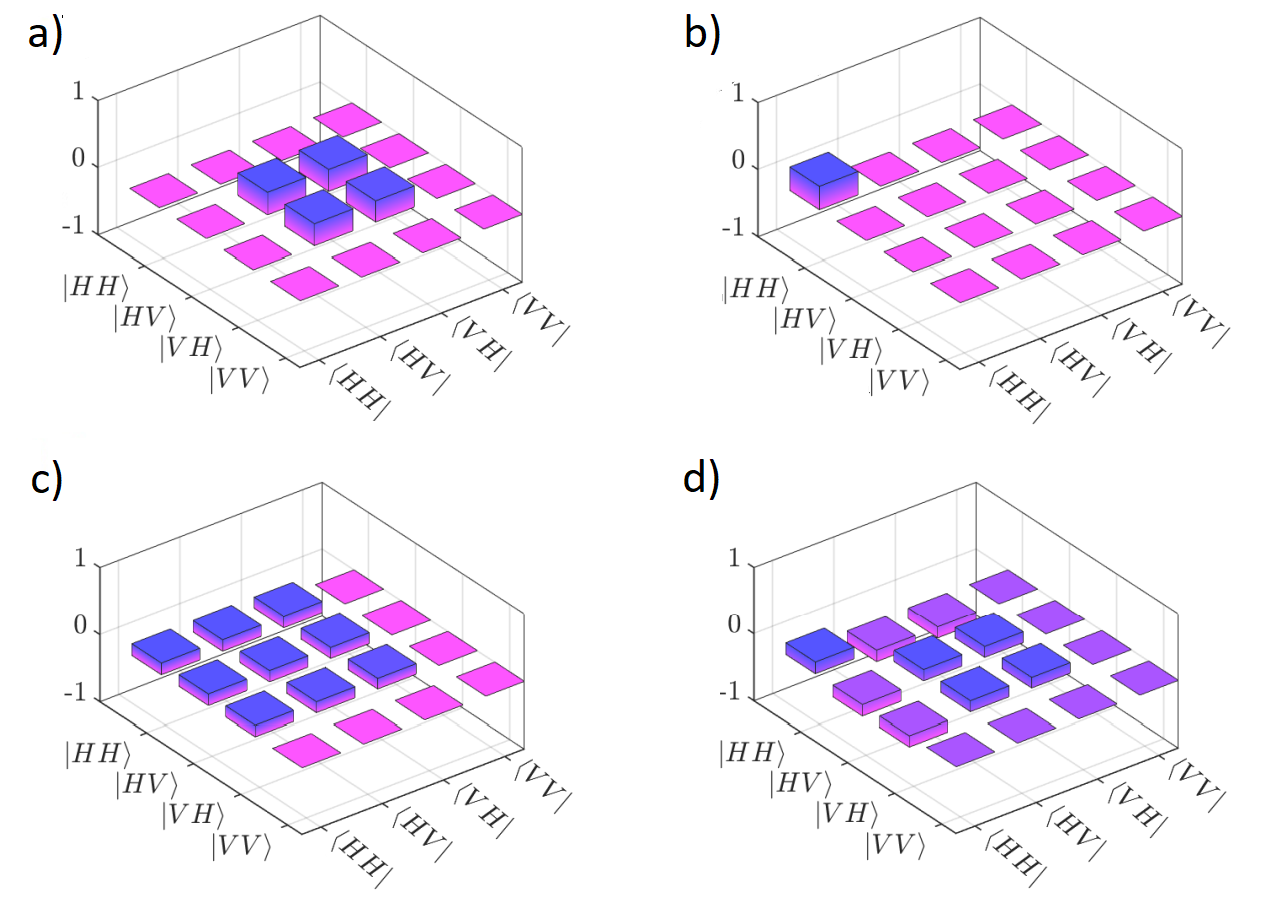}
    \caption{Results of the tomography for W-type states after Charlie's measurement of a) $\sigma_Z$ resulting 0, b)$\sigma_Z$ resulting 1, c) $\sigma_X$ resulting 0, and d) $\sigma_X$ resulting 1. The  results are the density matrix of the reduced state AB (spin-orbit state) multiplied by the corresponding weigh that depends on $\lambda_1$, $\lambda_2$,  and $\lambda_3$, in this case $\frac{1}{3}$.} 
    \label{rhoABW}
\end{figure}
\end{center}

The measurement circuit is the same used for GHZ states. The simulation of preparation and measurement for the 2-parameter W states using Jones Matrix formalism \cite{Jones:41} follows the same steps presented for GHZ states. Figure \ref{rhoABW} presents the measurement results for the usual W state where $\lambda_1 = \lambda_2= \lambda_3 =\frac{1}{\sqrt{3}}$. Fig.\ref{rhoABW}-a) and b) show the results of the tomography of the spin-orbit state for C's measurement of $\sigma_z$ resulting 0 and 1, respectively. Fig.\ref{rhoABW}-c) shows the results of the tomography of the spin-orbit state for C's measurement of $\sigma_x$ resulting 0 and Fig.\ref{rhoABW}-d) resulting 1. Here we also observed an excellent agreement with theoretical expectation. The fidelity are, respectively,  0.99, 0.99, 0.99, and 0.99. We obtain $E_A(\rho_{AB})- E_F(\rho_{AB})=0.1201$, that presents an error of $3\%$ with respect the theoretical expectation.
%
%
%

\section{Numerical results for more general states}

In this Section we perform an analysis considering more general states. To find the extremes of $\Bar{S}\left(\theta,\phi\right)$, a numerical search is conducted to determine the roots of the following set of Non-Linear Equations
\begin{align}
\label{eqn:partial_derivatives}
\begin{split}
 \frac{\partial}{\partial\theta}\Bar{S}\left(\theta,\phi\right) = 0,
\\
 \frac{\partial}{\partial\phi}\Bar{S}\left(\theta,\phi\right) = 0,
\end{split}
\end{align}
followed by an evaluation of the maximum and minimum values of $\Bar{S}\left(\theta,\phi\right)$ at the resulting stationary points. To solve the set~\eqref{eqn:partial_derivatives} for a wide range of states, especially those represented by dense matrices where the optimization problem posed by Eqs.~\eqref{obj_func_eoa} and~\eqref{obj_func_eof} is highly intricate, we employ the Globally Convergent Method described in Ref.~\cite{press1986numerical}. The method has proven to be suitable for our purposes, i.e., for calculating the EoA and EoF of arbitrary pure three-qubit states. 

In our analysis we considered five distinct types of states. Firstly,  for testing our numerical method we employed (i) $1\times 10^{3}$ 1-parameter GHZ type, as given by Eq.~\eqref{GHZ_type_state1}, and (ii) $1\times 10^{3}$ 2-parameter W type, as given by Eq.~\eqref{W_type_2_parameters1}, with randomly generated independent coefficients. Subsequently, we extended our analysis to (iii) $8\times10^{4}$ arbitrary pure tripartite states, which were generated randomly obeying the Haar measure. We also investigated (iv) $1\times 10^{2}$ arbitrary states obtained from the Generalized Schmidt Decomposition, as given by Eq.~\eqref{acin}, and (v) $1\times 10^{3}$ 3-parameter W states, as given by Eq.~\eqref{W_type_3_parameters1}, also with randomly generated independent coefficients.

\begin{figure}[h]
    \includegraphics[scale=0.22]{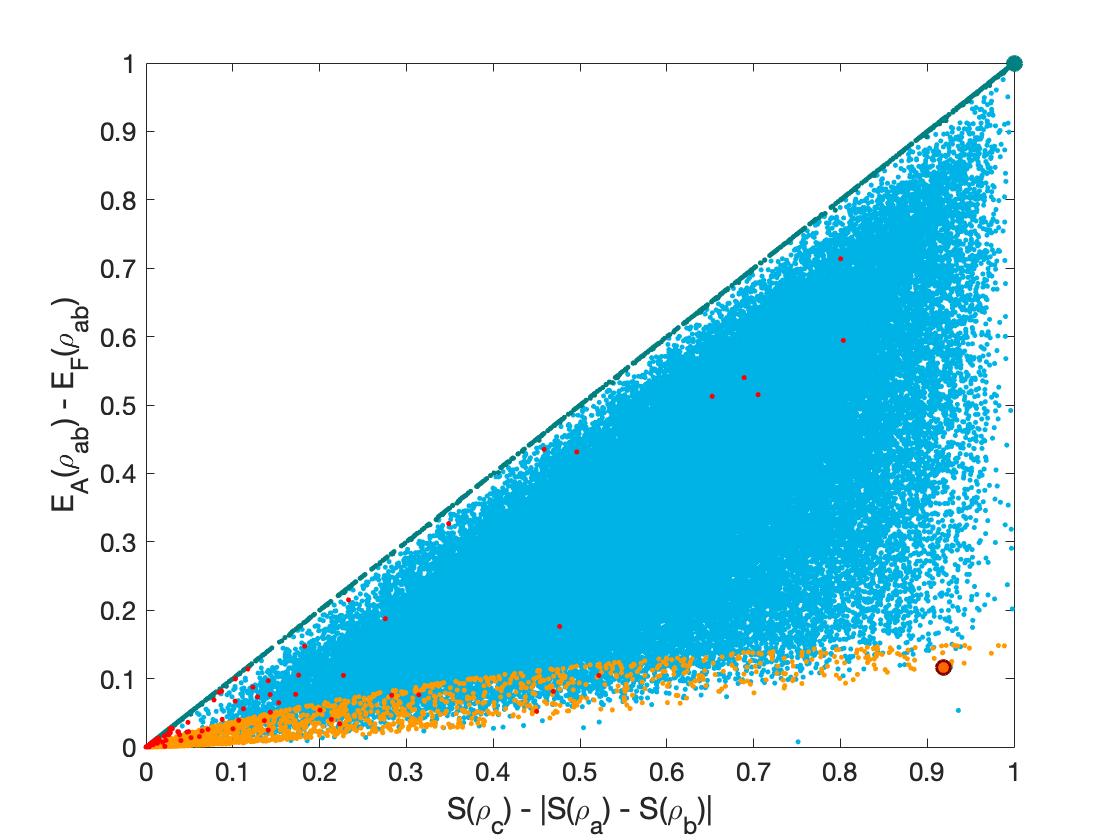}
    \caption{The (numerically solved) Activated Bipartite Entanglement plotted against its upper bound. The dots in light-blue represent a total of $8\times10^{4}$ randomly generated pure tripartite states satisfying the Haar measure. The dots in green over the straight line $E_{A}\left(\rho_ {AB}\right) - E_{F}\left(\rho_ {AB}\right) = S\left(\rho_{AB}\right) - \left\vert S\left(\rho_{A}\right) - S\left(\rho_{B}\right)\right\vert$ represent a total of $1\times10^{3}$ 1-parameter GHZ-states. The dots in orange represent a total of $2\times10^{3}$ W-states with half of them given by the 3-parameter W-type decomposition and the other half formed by the 2-parameter W-type decomposition. The dots in red represent a total of $1\times10^{2}$ arbitrary states given by Eq.~\eqref{acin}. The two big dots at coordinates (1,1) (green) and (0.11,0.9183) (orange) are, respectively, the ordinary GHZ and W states.}
    \label{fig:my_label}
\end{figure}

In Fig.~\ref{fig:my_label}, the light-blue dots are the $8\times10^{4}$ arbitrary pure tripartite states which obey the Haar measure. As we can observe, these dots spread almost all over the allowed area bellow the ABE's saturation line, $\Delta_{E}\left(\rho_{AB}\right) = \Delta S$, which in this case, is drawn in perfect agreement with the analytical solution by the $1\times 10^{3}$ green dots corresponding to the 1-parameter GHZ states.   
The delimited region bounded by the shadow of the W parametric surface projected on the $\Delta_{E}\left(\rho_{AB}\right)\times$ $\Delta S$ plane is home to the 2 and 3-parameter W states, illustrated in Fig.~\ref{fig:my_label} by $2\times 10^{3}$ orange dots (see Fig.~\ref{fig:exp}-bottom for an illustrative comparison). In our numerical analyses, the maximum value attained by the ABE for 3-parameter W states  is $\Delta_{E}\left(\rho_{AB}\right) \approx 0.15$, which is in accordance with the analytical results presented in Section~\ref{Exact_optimization_procedure} for the 2-parameter W states. Besides, both types are localised over the parametric surface represented in Fig.~\ref{fig:Surface} and we conclude that 2 and 3-parameter W states are completely equivalent in the ABE's parametric space. The sparse red dots scattered along the allowed area represent $1\times10^{2}$ 4-parameter arbitrary pure tripartite states given by Eq.~\eqref{acin} wich, for simplicity, we neglect the phase parameter: $\phi=0$. They encompass all types of pure tripartite states, i.e., separable, biseparable, GHZ and W states, much like the light-blue dots that represent arbitrary states obeying the Haar measure. 

To address the type of every state found in Fig.~\ref{fig:my_label}, our numerical analyses involve calculating the rank (denoted by $r(\rho_{k})$, $k=A, B, C$) of all reduced density matrices. Our findings reveal that any state for which $\Delta_{E}\left(\rho_{AB}\right)>0$ has $r(\rho_{A}) = r(\rho_{B}) = r(\rho_{C}) = 2$, which is a necessary condition for identifying genuine tripartite entanglement. This result is in perfect agreement with our previous analytical results, and we conclude that all the states in Fig.~\ref{fig:my_label} with $\Delta_{E}\left(\rho_{AB}\right)>0$ are necessarily GHZ or W states. To distinguish the GHZ-type from the W-type we have to look at the 3-tangle. Calculations show that the orange dots formed by 2 and 3-parameter W states possess 3-tangle \cite{kundu} equal to zero, which is indeed a characteristic of W states \cite{dur}. However, calculations also show that the arbitrary states obeying the Haar measure and the Generalized Schmidt Decomposition states with 4 independent parameters presenting null 3-tangle are located at the region bounded by the W-parametric surface. This observation supports the assumption that all the points plotted in Fig.~\ref{fig:my_label} that lie within the delimited orange region represent W states, including the light-blue dots and the red dots.  We highlight the ordinary W state with $\lambda_{0} = \lambda_{3} =1/\sqrt{3}$, depicted as the orange circle with coordinates ($\Delta_{E}\left(\rho^{W}_{AB}\right)$, $\Delta S^{W}$) = (0.11, 0.9183). The light-blue dots situated just above the W-region (including the green 1-parameter GHZ dots) exhibit a 3-tangle greater than zero, a unique feature solely attributed to the GHZ states. As a result, we can deduce that all states located above the W-region are GHZ states. The ordinary GHZ state given by $\lambda_{1} = 1/\sqrt{2}$ in Eq.~\eqref{GHZ_type_state1}, is represented in Fig.~\ref{fig:my_label} by the green circle with coordinates ($\Delta_{E}\left(\rho^{GHZ}_{AB}\right)$, $\Delta S^{GHZ}$) = (1,1). This state possesses the greatest value that ABE can assume. The dots at the origin correspond to all biseparable and separable states whose ranks are in agreement with the expected \cite{dur}: for biseparable states, one of them is equal to 1 and for completely separable states the three ranks are equal to 1.

\begin{figure}[h!]
    \includegraphics[scale=0.6]{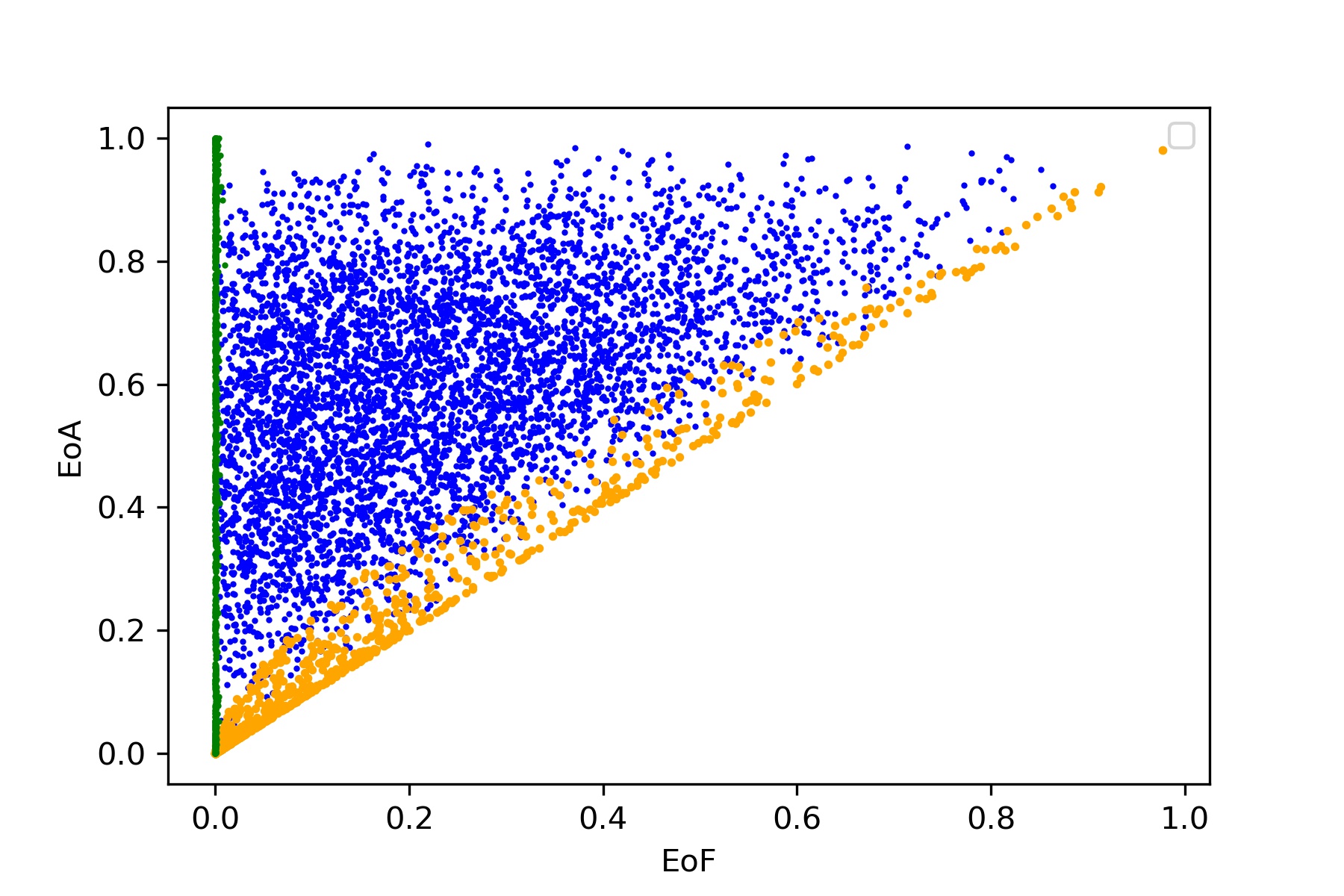}
    \caption{The (numerically solved) Scatter Plot of pairs of EoF x EoA. The dots in blue represent a total of $1\times10^{3}$ randomly generated pure tripartite states satisfying the Haar measure. The dots in green over the straight EoF = 0 represent a total of $1\times10^{3}$ 1-parameter GHZ-states. The dots in orange represent a total of $2\times10^{3}$ W-states with half of them given by the 3-parameter W-type decomposition and the other half formed by the 2-parameter W-type decomposition.}
    \label{fig:fig9}
\end{figure}

Besides, the investigation of entanglement within GHZ and W states presented in this section provides new insights into the quantum entanglement landscape. Our study demonstrates that contrary to the prevailing assumptions, arbitrary GHZ states given by Eq.~\eqref{acin} and W states exhibit significant and measurable EoF values under specific conditions. These findings challenge the traditional perception that GHZ states always exhibits null EoF. In Fig.~\eqref{fig:fig9} we illustrate how these states can indeed sustain substantial entanglement levels as quantified by EoF by means of a EoF x EoA scatter plot. As we can see, the bulk of dots in blue represent GHZ states of the form given by Eq.~\eqref{acin}, also a total of $1\times10^{3}$ randomly generated pure tripartite states satisfying the Haar measure. The dots in green over the straight EoF = 0 represent a total of $1\times10^{3}$ 1-parameter GHZ-states. The dots in orange represent a total of $2\times10^{3}$ W-states with half of them given by the 3-parameter W-type decomposition and the other half formed by the 2-parameter W-type decomposition This revelation not only augments our understanding of tripartite entanglement but also suggests potential areas for further experimental and theoretical exploration. Our results underline the nuanced nature of entanglement in these canonical states and pave the way for utilizing their complex entanglement properties in advanced quantum technologies and computational schemes.

\section{Conclusions}

We have developed a method for detecting genuine tripartite entanglement in pure 3-qubit states using the Activated Bipartite Entanglement (ABE). Our approach indicates that when such states exhibit genuine tripartite entanglement, the ABE is necessarily greater than zero; otherwise, it equals zero. This method can also differentiate between genuine tripartite entangled states belonging to the GHZ class and those belonging to the W class. We solved the optimization problem encountered in the expressions of EoA and EoF analytically for 1-parameter GHZ states and 2-parameter W states, and numerically for more general 3-qubit states. We demonstrate that the 1-parameter GHZ state saturates the upper bound of ABE, indicating its exceptional value as a resource for entanglement that can be localised in $AB$ with the assistance of $C$. We demonstrate that all W states, regardless of whether they are 3-parameter or 2-parameter W types, must lie on a particular parametric surface that characterizes these states. By redefining the upper limit of ABE for W states, we prove that any state with an ABE value greater than or equal to 0.15 must be a GHZ state. While our method for distinguishing between GHZ and W states may be more challenging to implement due to the need to solve an optimization problem, it offers the advantage of providing a visual representation of their distribution in ABE's parametric space. In contrast to the rank and 3-tangle measurement method, which lack this perspective, our approach offers unique insights. In fact, each method has its own strengths and complements the others. Moreover, we also presented an experimental proposal for measuring ABE and implementing our method, by using linear circuits to prepare tripartite states codifying qubits in internal degrees of freedom of a single photon. The optical circuits was designed to prepare GHZ and W states and perform all measurements. Simulations of the optical circuits show an excellent agreement with theory.

More significant, the present work has drawn an important conclusion that GHZ states represent a superior and more efficient resource when it comes to addressing the issue of localizing entanglement between two separate components with the assistance of a distant third component. 

We should remark that, as we have only considered projective operations on $C$, it resulted in the ABE being defined in terms of the EoA and EoF. We could extend the formalism to general POVM, and it would give similar results in terms of the Localizable Entanglement \cite{PhysRevA.71.042306}, instead. However, this approach would be more difficult to calculate and to experimentally implement. Besides, it would not bring any additional insight.

Future investigation can be performed for the extension of ABE to mixed state. Considering experimental proposal, recently it was proposed a linear optical circuit to prepare tripartite mixed states using degree of freedom of light of light \cite{MixedGHZ}. Then, we have the experimental platform to extend investigation of ABE for mixed state.

We would like to thank financial support from the Brazilian funding agencies Conselho Nacional de Desenvolvimento Cient\'{\i}fico e Tecnol\'ogico (CNPq),
through the Brazilian National Institute for Science and Technology of Quantum Information (INCT-IQ),
Funda\c{c}\~ao Carlos Chagas Filho de Amparo \`a Pesquisa do Estado do Rio de Janeiro (FAPERJ), 
Coordena\c{c}\~ao de Aperfei\c{c}oamento de Pessoal de N\'{\i}vel Superior (CAPES). L.G.E.A. would like to thank Felipe F. Fanchini for the computational support provided.

\bibliographystyle{apsrev4-1}
\bibliography{main}

\end{document}